\pdfoutput=1
\documentclass[
reprint,
superscriptaddress,
 amsmath,amssymb,
 aps,
 prd,
]{revtex4-2}

\usepackage[table,svgnames,dvipsnames]{xcolor}
\definecolor{Blue}{HTML}{1c1e94}
\usepackage[colorlinks=true,citecolor=Blue,linkcolor=Blue,urlcolor=Blue, backref=false,pdfborder={0 0 0}]{hyperref}
\usepackage[normalem]{ulem}
\usepackage{aas_macros}
\usepackage{graphicx}% Include figure files
\usepackage{dcolumn}% Align table columns on decimal point
\usepackage{bm}% bold math
\usepackage[mathlines]{lineno}% Enable numbering of text and display math
\usepackage{orcidlink}
\usepackage{cases}% for dealing with mathematics,
\usepackage{booktabs}
\usepackage{comment}
\usepackage{multirow}
\usepackage{makecell}
\usepackage{siunitx}
\usepackage{booktabs}
\graphicspath{{figs/}}

\usepackage{fontawesome5}
\makeatletter
\newcommand{\github}[1]{%
   \href{#1}{\faGithub}%
}
\makeatother

\def\be{\begin{equation}}
\def\ee{\end{equation}}

\def\ba#1\ea{\begin{align*}#1\end{align*}}
\newcommand\numberthis{\addtocounter{equation}{1}\tag{\theequation}}

\newcommand{\code}[1]{{\texttt{#1}}}
\renewcommand{\emph}[1]{\textit{#1}}

\usepackage[nameinlink,noabbrev]{cleveref}
\crefname{equation}{Eq.}{Eqs.}
\crefname{section}{Section}{Sections}
\crefname{figure}{Fig.}{Figs.}
\crefname{table}{Table}{Tables}
\crefname{appendix}{Appendix}{Appendices}
\Crefname{figure}{Figure}{Figures}
\Crefname{equation}{Equation}{Equations}
\Crefname{section}{Section}{Sections}
\Crefname{table}{Table}{Tables}

\newcommand{\<}{\left\langle}
\renewcommand{\>}{\right\rangle}
\newcommand{\convD}{\frac{\mathrm{D}}{\mathrm{D}\tau}}

\renewcommand{\L}{\mathcal{L}}
\renewcommand{\P}{\mathcal{P}}

\newcommand{\R}{\mathcal{R}}
\newcommand{\U}{\mathcal{U}}

\newcommand{\N}{\mathcal{N}}

\newcommand{\cov}{\mathsf{C}}

\newcommand{\fbi}{\mathrm{FBI}}
\newcommand{\fid}{\mathrm{fid.}}
\newcommand{\true}{\mathrm{true}}
\newcommand{\obs}{\mathrm{obs.}}
\newcommand{\eff}{\mathrm{eff.}}
\newcommand{\expl}{\mathrm{expl.}}
\newcommand{\impl}{\mathrm{impl.}}
\newcommand{\unmarg}{\mathrm{unmarg.}}

\renewcommand{\max}{\mathrm{max}}

\newcommand{\kmax}{k_\max}

\newcommand{\eps}{\epsilon}
\newcommand{\sigmaEps}{\sigma_\eps}
\DeclareMathOperator{\tr}{tr}

\newcommand{\PplusB}{\mathrm{P\!+\!B}}
\newcommand{\Otd}{O_{\mathrm{td}}}
\newcommand{\Del}{\mathcal{D}}

\newcommand{\Om}{\Omega_m}

\newcommand{\shat}{\hat{s}}
\newcommand{\shattrue}{\hat{s}_{\true}}
\renewcommand{\d}{\delta}
\newcommand{\dg}{\delta_g}

\newcommand{\Dplus}{D^{(1)}_+}
\newcommand{\Plin}{P_{\mathrm{L}}}

\newcommand{\dlin}{\delta^{(1)}}

\newcommand{\Nsim}{N_{\mathrm{sim}}}
\newcommand{\Nmode}{N_{\mathrm{mode}}}
\newcommand{\Ngrid}{N_{\mathrm{grid}}}
\newcommand{\NgridEul}{N^{\mathrm{Eul}}_{\mathrm{grid}}}

\newcommand{\NgridNUFFT}{N^{\rm NUFFT}_{\mathrm{grid}}}
\newcommand{\Nbin}{N_{\mathrm{bin}}}
\newcommand{\Ntriangle}{N_{\mathrm{triangle}}}

\newcommand{\Mpch}{\,h^{-1}\mathrm{Mpc}}

\newcommand{\iMpch}{\,h\,\mathrm{Mpc}^{-1}}

\newcommand{\Msunh}{\,h^{-1} M_\odot}

\newcommand{\vx}{\bm{x}}
\newcommand{\vk}{\bm{k}}
\newcommand{\vq}{\bm{q}}

\begin{document}

\title{How much information can be extracted from galaxy clustering at the field level?}

\author{Nhat-Minh Nguyen\orcidlink{0000-0002-2542-7233}}
\email{nguyenmn@umich.edu}
\affiliation{Leinweber Center for Theoretical Physics, University of Michigan, 450 Church St, Ann Arbor, MI 48109-1040}
\affiliation{Department of Physics, College of Literature, Science and the Arts, University of Michigan, 450 Church St, Ann Arbor, MI 48109-1040}
\author{Fabian Schmidt\orcidlink{0000-0002-6807-7464}} \email{fabians@MPA-Garching.MPG.DE}
\affiliation{Max–Planck–Institut für Astrophysik, Karl–Schwarzschild–Straße 1, 85748 Garching, Germany}
\author{Beatriz Tucci\orcidlink{0000-0003-2971-2071}} \email{tucci@MPA-Garching.MPG.DE}
\affiliation{Max–Planck–Institut für Astrophysik, Karl–Schwarzschild–Straße 1, 85748 Garching, Germany}
\author{\\Martin Reinecke}
\affiliation{Max–Planck–Institut für Astrophysik, Karl–Schwarzschild–Straße 1, 85748 Garching, Germany}
\author{Andrija Kosti\'{c}\orcidlink{0000-0002-8219-0025}}
\affiliation{Max–Planck–Institut für Astrophysik, Karl–Schwarzschild–Straße 1, 85748 Garching, Germany}

\date{\today}

\begin{abstract}
We present optimal Bayesian field-level cosmological constraints from nonlinear tracers of cosmic large-scale structure, specifically the amplitude $\sigma_8$ of linear matter fluctuations inferred from rest-frame simulated dark matter halos in a comoving volume of $8\,(h^{-1}\mathrm{Gpc})^3$.
Our constraint on $\sigma_8$ is entirely due to nonlinear information, and obtained by explicitly sampling the initial conditions along with tracer bias and noise parameters via a Lagrangian EFT-based forward model, \code{LEFTfield}.
The comparison with a simulation-based inference of the power spectrum and bispectrum---likewise using the \code{LEFTfield} forward model---shows that, when including precisely the same modes of the same data up to $\kmax = 0.10\iMpch$ ($0.12\iMpch$), the field-level approach yields a factor of 3.5 (5.2) improvement on the $\sigma_8$ constraint, going from 20.0\% to 5.7\% (17.0\% to 3.3\%).
This study provides direct insights into cosmological information encoded in galaxy clustering beyond low-order $n$-point functions.
\end{abstract}

\maketitle

\textit{Introduction.}---Cosmological surveys of large-scale structure (LSS) have significantly advanced from measuring angular clustering of galaxies identified on photographic plates to mapping three-dimensional clustering with spectroscopy and optical fibers positioned by robots.
Meanwhile, statistical methods to analyze galaxy clustering and other biased tracers of LSS still largely rely on modeling two- and three-point correlation functions.
This leaves open the question:
\begin{center}
\emph{How much cosmological information can be robustly extracted from galaxies and LSS tracers?}
\end{center}
In this Letter, we address this question by comparing constraints on the amplitude of linear matter fluctuations, $\sigma_8$ [\cref{eq:sigma8_def}] obtained with (1) the full field-level statistics, i.e. the entire three-dimensional tracer field, and (2) the combination of power spectrum plus bispectrum ($\PplusB$) summary statistics.
In both setups, we analyze the same data using the same data model based on the Effective Field Theory (EFT) of LSS \cite{Baumann:2010tm,Carrasco:2012cv,Carroll:2013oxa}.

Constraints on $\sigma_8$, derived from LSS surveys and cosmic microwave background (CMB) experiments, quantify the linear growth of structure since the primordial universe, which in turn probes Dark Energy and modifications to General Relativity. Recently, $\sigma_8$, or the parameter combination $S_8=\sigma_8\sqrt{\Om/0.3}$, has been under investigation since tensions arise between CMB constraints \cite{Planck:2018vyg,ACT:2020gnv,SPT-3G:2022hvq,ACT:2023kun} and galaxy weak lensing constraints \cite{Heymans:2020gsg,DES:2021wwk,Kilo-DegreeSurvey:2023gfr,Sugiyama:2023fzm,Lange:2023khv}, within the standard model $\Lambda$CDM.
While CMB measures matter fluctuations in the early universe on the largest observable scales, galaxy weak lensing quantifies fluctuations in the late universe on smaller scales, making it hard to disentangle the physical origin of the tension: Is it a tension between physics of the early and late universe \cite{Poulin:2022sgp,Nguyen:2023fip,Lin:2023uux} or between that of large and small scales \cite{Amon:2022azi,Rogers:2023ezo,Preston:2023uup}, or both?
Separate $\sigma_8$ constraints from galaxy clustering---which alone probes growth in the late universe on large-to-intermediate scales---therefore provide an additional crucial data point to dissect this tension.

\textit{$\sigma_8$ constraint from galaxy clustering.}---The initial matter fluctuations out of which LSS formed closely follow a zero-mean multivariate Gaussian distribution $\N(0,\cov)$ \cite{Planck:2019evm}.
At late times, the linear growth of matter fluctuations $\dlin(\vk,z)$  in Fourier space can be decomposed into the scale-dependent transfer function $T(k)$ and the redshift-dependent growth function $\Dplus(z)$, such that $\dlin(\vk,z)=(2/5)(k^2/\Om H_0^2)\Dplus(z)T(k)\R(k)$ for each Fourier mode $\vk$, where $\R(\vk)$ is the primordial curvature perturbation, while $\Om$ and $H_0$ are the matter density parameter and the Hubble constant, respectively.
A proxy of growth up to today can be defined through the variance of linear matter fluctuations at $z=0$:
\be
\sigma_8^2 \equiv \< \left(\dlin_R(\vx)\right)^2\>
= \int_0^\infty \frac{dk\,k^2}{2\pi^2}\Plin(k,z=0)W^2_R(k),
\label{eq:sigma8_def}
\ee
where $\Plin$ is the linear power spectrum and $W_R$ is a spherical window function of radius $R=8\Mpch$.

Constraining $\sigma_8$ from galaxy clustering is complicated by the fact that galaxies are discrete, biased tracers of the underlying matter field.
At leading order in perturbations, the observed galaxy power spectrum is related to the linear matter power spectrum through $P_{\dg^\obs }(k,z)=b_\delta^2(z)\Plin(k,z)+P_\eps(z)$ where $b_\delta$ is the linear galaxy bias coefficient and $P_\eps$ is the amplitude of the galaxy stochastic contribution, both of which encapsulate the a priori unknown physics of galaxy formation. At this order, there is thus a perfect degeneracy between $b_\delta$ and $\sigma_8$, which can only be broken by considering nonlinear clustering or redshift-space distortions (RSD). However, at leading order, RSD exhibits a further degeneracy in that it only constrains the product $f \sigma_8$, with the growth rate $f = d\ln D_+/d\ln a$.

How does nonlinear clustering information help with the $\sigma_8-b_\delta$ degeneracy?
Beyond linear order, both the matter $\d$ and galaxy density $\d_g$ contain advection contributions involving the Lagrangian displacement field.
Einstein's equivalence principle ensures that matter and galaxies co-move on large scales \cite{Desjacques:2016bnm}. The latter implies that the displacement is the same for galaxies and matter, which means that the advection contribution to $\d_g$ can be uniquely predicted. This direct consequence of the equivalence principle therefore breaks the degeneracy between $b_\d$ and $\sigma_8$.

This work provides the first proof that (robust) constraints on $\sigma_8$ can be extracted from nonlinear clustering of dark-matter halos in N-body simulations---directly at the field level. Field-level inference explicitly samples the full posterior of $[\sigma_8,\dlin]$ given the data, and is hence guaranteed to be optimal within the assumed data model \footnote{We note that some previous studies in the context of cosmological parameter inference, e.g. \cite{Lemos:2023myd,Hahn:2023udg}, used the term ``field-level inference'' for a different analysis. As their inference pipeline relied on neural network data compression, there is in fact no guarantee that this neural compression is lossless and the encoded summary statistics capture all relevant cosmological information}.

Our analysis extends previous works in several ways. Firstly, \cite{Schmidt:2020ovm,Schmidt:2020tao} demonstrated the ability of the EFTofLSS framework to predict clustering in N-body simulations at the field level to percent-level precisions, conditioned on the true initial conditions $\shat_{\mathrm{true}}$ of the simulations. The initial conditions $\shat$ of the universe, however, are not direct observables. Here, we jointly infer and marginalize over the unknown initial conditions.
Thus, our $\sigma_8$ constraints are much more representative of those in a real-world scenario.

 Secondly, in this work, we study N-body halos---representative of the true observable in real galaxy surveys---in contrast to previous studies that analyzed either mock data drawn from the same forward model \cite{Ramanah:2018eed,Andrews:2022nvv,Kostic:2022vok} or idealized, unbiased tracers, namely dark-matter particles \cite{Seljak:2017rmr,Bayer:2023rmj,Mudur:2023smm}. Since we cannot (yet) simulate realistic galaxy catalogs in a cosmological volume, model misspecification is inherent in any data model and analysis, and the robustness of the results against such misspecification must be validated.
In fact, this is likely to be the key challenge for field-level inference \cite{Nguyen:2020hxe,Villanueva-Domingo:2022rvn}.
Recovering unbiased constraints on $\sigma_8$ in this work thus is a stringent test of the robustness of the EFTofLSS field-level forward model and inference framework. Given that our forward model is agnostic to details of the tracer---and based only on the equivalence principle---we expect that our results generalize to actual galaxies as well (see \cite{2021JCAP...08..029B,Beyond2pt:preprint} for field-level results on simulated galaxies).

Lastly, this is the first time a consistent comparison between field-level and summary statistics has been conducted on the basis of exactly the same galaxy clustering data and model.

\textit{Data.}---Our primary halo sample \code{SNG} consists of main halos in the $\log_{10} M_{200m}=12.5-14.0\Msunh$ mass range, identified with the \code{ROCKSTAR} halo finder \cite{Behroozi:2011ju} at redshift $z=0.50$ in an N-body, gravity-only simulation (mean comoving number density $\bar n= 1.3\cdot10^{-3}(\Mpch)^{-3}$). The simulation assumes a $\Lambda$CDM cosmology with $\sigma_{8,\true}=0.850$, encompasses a comoving volume $L^3=(2000\Mpch)^3$ and contains $N_{\mathrm{particle}}=1536^3$ particles of mass $M_{\mathrm{particle}} = 1.8\times10^{11}\Msunh$ \cite{Schmidt:2020tao}.

We additionally analyze the \code{Uchuu} halo sample, consisting of main halos in the $\log_{10} M_{200m}=12.0-13.5\Msunh$ mass range, identified with \code{ROCKSTAR} at redshift $z=1.03$ ($\bar n= 3.6\cdot10^{-3}(\Mpch)^{-3}$) in the \code{Uchuu} simulation \cite{Ishiyama:2020vao}, which assumes $\sigma_{8,\true}=0.816$. This simulation spans the same volume $L^3=(2000\Mpch)^3$ while offering a much higher mass resolution, $N_{\mathrm{particle}}=12800^3$ particles of mass $M_{\mathrm{particle}} = 3.27\times10^{8}\Msunh$.

\textit{Field-level EFT of biased tracers.}---The EFTofLSS provides a perturbative framework within which the galaxy density field $\dg$ can be systematically expanded, order by order, in perturbations as
\be
\dg(\vk,z)=\sum_O b_O(z) O(\vk,z) + \eps(\vk,z).
\label{eq:deltag_def}
\ee
The first term on the r.h.s. of \cref{eq:deltag_def} includes deterministic contributions, such as $b_\d(z) \d(\vk,z)$,  and describes the biased nature of galaxies as tracers of the underlying matter field.

At each given order, there is a finite number of galaxy bias operators $O$, each associated with a coefficient $b_O$.
The operators $O$ are constructed out of local gravitational observables, obeying the equivalence principle.
Further details on the galaxy bias expansion, operators and coefficients can be found in the Supplementary Material.

The second term on the r.h.s. of \cref{eq:deltag_def} is the stochastic contribution encapsulating the random nature of small-scale fluctuations and the discrete nature of galaxies. These are described by the noise field $\eps$, which to leading order is Gaussian with RMS
\be
\sigma_\eps(k) = \sigma_{\eps,0}\left[1 + \sigma_{\eps,k^2}k^2\right],
\label{eq:sigmaEps_def}
\ee
where the subleading contribution $\propto k^2$ captures the non-locality of galaxy formation.
\begin{figure*}[t]
        \centering
            \includegraphics[width=0.8\linewidth]{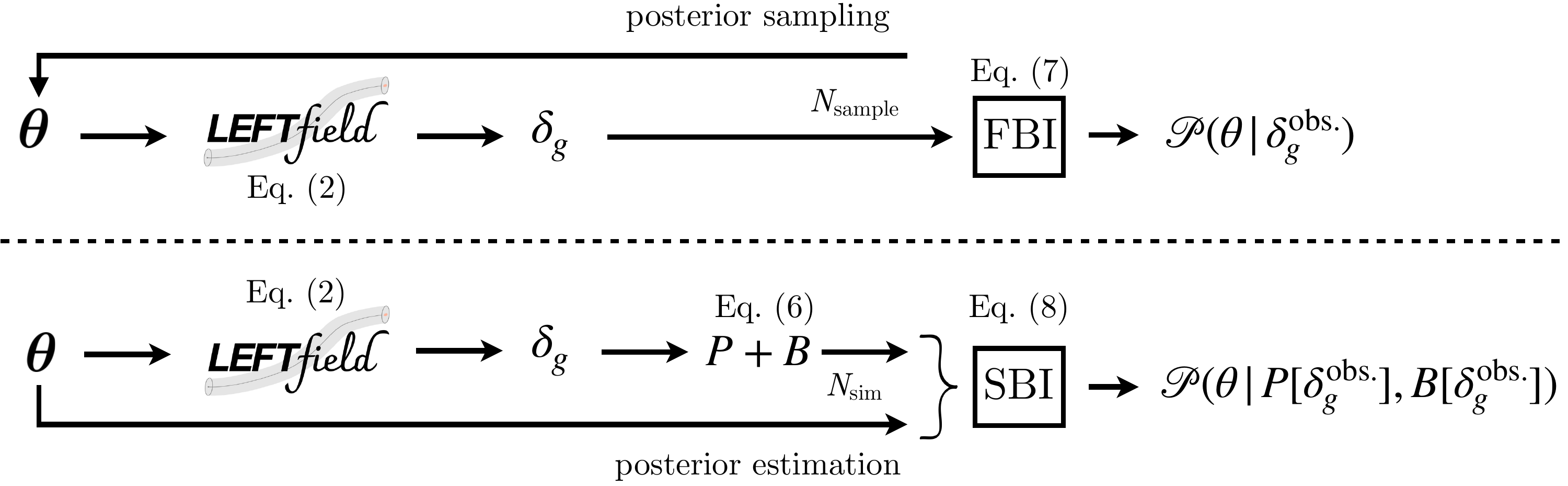}
            \caption{Diagram of the two inference methods, FBI and SBI $\PplusB$. Both pipelines share the same forward model \code{LEFTfield}.}
            \label{fig:FBI_SBI-PplusB_diagram}
 \end{figure*}

\medskip
\textit{Field-level forward model.}---We forward model the bias fields $O$ in \cref{eq:deltag_def} starting from Gaussian initial conditions parametrized via $\shat\sim\N(0,1)$, discretized on a grid of size $\Ngrid$, hence
\be
\dlin(\vk,z)=\alpha\,\left[\frac{\Ngrid^3}{L^3}\Plin(k,z)\right]^{1/2}\,\shat(\vk),
\label{eq:deltalin_shat}
\ee
where $L$ is the simulation box side length, and $\alpha=\sigma_8/\sigma_{8,\true}$ is the amplitude rescaling parameter. $\alpha=1$ indicates an unbiased inference for either halo sample.

Our forward model employs the Lagrangian, EFT-based forward model of cosmological density fields, \code{LEFTfield} \cite{Schmidt:2020ovm}.
Following the EFT principle, \code{LEFTfield} evolves all cosmological (plus auxiliary) fields up to a finite EFT cutoff $\Lambda$.
We choose $\Lambda$ to be a sharp-$k$ filter that strictly filters out all Fourier modes above the cutoff \cite{Carroll:2013oxa,Schmidt:2018bkr,Schmidt:2020viy}. Specifically, we implement a cubic sharp-$k$ filter via a Fourier grid reduction \cite{Schmidt:2020tao}.
Crucially, \code{LEFTfield} computes both $O=O(\shat)$ and $\partial O/\partial\shat$, the latter of which proves useful for gradient-based sampling and field-level inference.
We refer to Supplementary Material and \cite{Schmidt:2020ovm,Schmidt:2020viy,Kostic:2022vok,Tucci:2023bag} for \code{LEFTfield} implementation and validation.
Here, the new developments with respect to \cite{Schmidt:2020ovm,Schmidt:2020viy,Kostic:2022vok,Tucci:2023bag} are:
(1) a third-order model for galaxy bias, improving accuracy relative to the previous (second-order) treatment;
(2) a non-uniform Fast Fourier Transform (NUFFT \cite{NUFFT}) for grid assignment, enhancing numerical convergence and efficiency relative to previous assignment schemes; and (3) a change to $\kmax = \Lambda/1.2$, effectively reducing the analysis cutoff scale $\kmax$ relative to the initial conditions cutoff $\Lambda$, hence mitigating higher-derivative contributions.

\medskip
\textit{Inference method I: FBI with explicit likelihood.}---In the field-level Bayesian inference (FBI) pipeline, we evaluate and sample from an explicit field-level likelihood $\L_\fbi^\expl$, depicted in the top row of \cref{fig:FBI_SBI-PplusB_diagram}.

Following \cite{Cabass:2020nwf}, our fiducial analyses assumes Gaussianity of galaxy stochasticity and analytically marginalizes over $\eps$.
This leads to a Gaussian likelihood of the following form for an observed and filtered galaxy field $\d^\obs_{g}$ \cite{Schmidt:2018bkr,Cabass:2019lqx}:
\ba
&\L_\fbi^\expl\left(\d^\obs_{g}  \Big|\shat, \alpha, \{b_O\}, \{\sigmaEps\}\right) =
-\frac{1}{2}\sum_{\vk > 0}^{|\vk| < \kmax}\\
&\Big[\ln{2\pi\sigmaEps^2(k)}
+\frac{1}{\sigmaEps^2(k)}
\Big\lvert
\d^\obs_{g}(\vk) - \sum_O b_O O[\alpha, \shat](\vk)
\Big\rvert^2
\Big]\,\,.\numberthis\label{eq:FBI_likelihood}
\ea
The $\sum_{\vk > 0}^{|\vk| < \kmax}$ amounts to a spherical sharp-$k$ filter which only includes Fourier modes $\vk$ up to $\kmax$, the cutoff scale of our analyses while also excludes the $\vk=0$ mode.
We expand $\sum_Ob_OO$ to third order in the galaxy bias operators $O$ and further analytically marginalize over the bias coefficients $\{b_O\}$ assuming weakly informative Gaussian priors (see Supplementary Material).

The final explicit FBI parameter space consists of $[\shat,\alpha,\{\sigmaEps\}]$.
The element $\shat$ is a three-dimensional grid of size $[\Ngrid\times\Ngrid\times\Ngrid]$ containing $\Nmode=\Ngrid^3$ modes of initial density fluctuations.
To explore this high-dimensional posterior, following \cite{Jasche:2018oym,Kostic:2022vok}, we employ two MCMC sampling methods: Hamiltonian Monte Carlo (HMC) \cite{Neal:HMC} for $\shat$---leveraging the differentiability of \code{LEFTfield} forward models---and slice sampling \cite{Neal:slice} for $[\alpha,\{\sigmaEps\}]$.

\medskip

\textit{Inference method II: SBI $\PplusB$ with implicit likelihood.}---Implicit-likelihood or simulation-based inference (SBI) directly learns the posterior from simulated training data without assuming any analytical form for the likelihood of the data vector \cite{Cranmer:2020}.
Our SBI $\PplusB$ pipeline is depicted in the bottom row of \cref{fig:FBI_SBI-PplusB_diagram}, where we closely follow the procedure detailed in \cite{Tucci:2023bag}.
We first draw the parameters $\theta\equiv[\alpha,\{b_O\},\{\sigmaEps\}]$ from their priors and simulate the galaxy fields $\d_g$ via \cref{eq:deltag_def} \cref{eq:FBI_likelihood} with \code{LEFTfield}.
We then measure the power spectrum $P$ and bispectrum $B$ on each simulated data realization,
\begin{subequations}
\begin{align}
    &\left\langle \d_g(\mathbf{k}) \d_g(\mathbf{k'}) \right\rangle = P(k) (2\pi)^3 \d_D(\mathbf{k}+\mathbf{k'}), \\
    &\begin{aligned}
        &\left\langle \d_g(\mathbf{k}_1) \d_g(\mathbf{k}_2) \d_g(\mathbf{k}_3) \right\rangle = \\
        &\qquad\:\;B(k_1,k_2,k_3)(2\pi)^3\delta_D(\mathbf{k}_1+\mathbf{k}_2+\mathbf{k}_3),
    \end{aligned}
\end{align}
\end{subequations}
following \cite{Scoccimarro:1997st} (see also Eqs.~(2.15--2.17) of \cite{Tucci:2023bag}).
The SBI $\PplusB$ data vector contains $\Nbin+\Ntriangle$ elements up to the same $\kmax$ used in the FBI analysis, with $\Nbin$ linear $k$-bins for the power spectrum and $\Ntriangle$ triangle $k$-configurations for the bispectrum. We choose a $k$ bin width of $\Delta k=2k_f$, where $k_f\equiv2\pi L^{-1}$ is the fundamental frequency.
The $N_{\mathrm{sim}}$ samples, drawn from the joint distribution $\left\{\theta,\  P[\d_{g}(\theta)],B[\d_{g}(\theta)] \right\}$ this way, form the SBI training set.
We use neural posterior estimation (NPE) \cite{Greenberg:2019} with masked autoregressive flows \cite{Papamakarios:2017} from the \code{sbi} package \cite{Tejero-Cantero:2020} (see Supplementary Material).

After training, we sample the estimated posterior $\P_{\PplusB}$, conditioned on the power spectrum plus bispectrum measured on the ``observed'' data 
$\left[P[\d^\obs_{g}], B[\d^\obs_{g}]\right]$, [\cref{eq:PplusB_posterior}].
We employ simulation-based calibration (SBC) \cite{Talts:2018} and convergence tests to validate the SBI posteriors (see Supplementary Material).
We note that the forward model employed here assumes Gaussian noise [\cref{eq:FBI_likelihood}]. Thus our bispectrum model does not contain a contribution from non-Gaussian (skewed) noise, or from a density-dependent noise variance.
In the Supplementary Material, we compare our fiducial SBI $\PplusB$ analysis with a variant that includes both additional stochastic contributions, but employs a restricted bias parameter set (see Supplementary Material). This variant matches current standard $\PplusB$ analyses \cite{Philcox:2021kcw,DAmico:2022osl}.
We find broad consistency between both SBI $\PplusB$ analyses.

\textit{Inference summary:}---Explicitly, our target posteriors are
        \begin{widetext}
            \ba
                \P_{\fbi}\left(\alpha, \{b_O\}, \{\sigmaEps\} \Big| \d^\obs_{g}  \right) 
                &\propto \int\mathcal{D}\shat\,\P\left(\shat\right)\,\L_\fbi^\expl\left(\d^\obs_{g}  \Big|\shat, \alpha, \{b_O\}, \{\sigmaEps\}\right)\,\P\left(\alpha,\{b_O\},\{\sigmaEps\}\right),\numberthis\label{eq:FBI_posterior}\\
                \P_{\PplusB}\left(\alpha, \{b_O\}, \{\sigmaEps\} \Big| P[\d^\obs_{g}], B[\d^\obs_{g}] \right)
                &\propto \L_\PplusB^\impl\left(P[\d^\obs_{g}], B[\d^\obs_{g}] \Big|\alpha, \{b_O\}, \{\sigmaEps\}\right)\,\P\left(\alpha,\{b_O\},\{\sigmaEps\}\right),\numberthis\label{eq:PplusB_posterior}
            \ea
        \end{widetext}
for FBI [\cref{eq:FBI_posterior}] and SBI $\PplusB$ [\cref{eq:PplusB_posterior}], where $\{b_O\}$ consists of all bias parameters up to third order. We explicitly list the bias parameters $\{b_O\}$ and the priors $\P\left(\alpha,\{b_O\},\{\sigmaEps\}\right)$ in the Supplementary Material.

\begin{figure}[th]
%   \centering
   \includegraphics[width=0.8\linewidth]{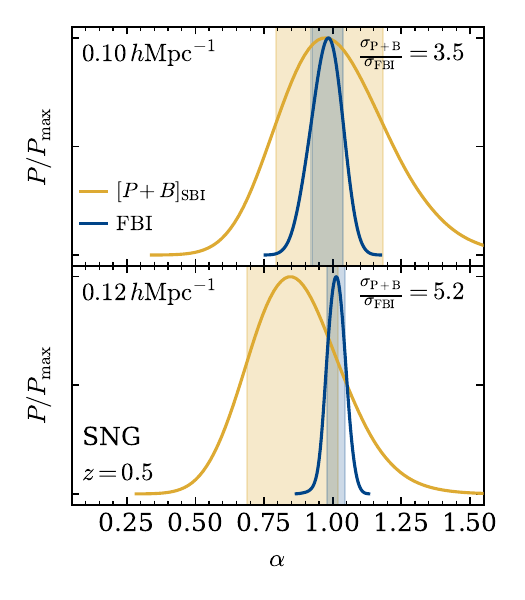}
   \caption{Constraints on $\alpha=\sigma_8/\sigma_{8,\true}$, from the \code{SNG} sample (see text), at $\kmax=[0.10,0.12]\iMpch$. Vertical bands indicate the 68\% limits of the posteriors. The ratios of the 1-$\sigma$ constraints between FBI (blue) and SBI $\PplusB$ (yellow) are shown in the upper right corners.}
   \label{fig:alpha_1D-constraint_FBI_vs_P+B_SNGz05_2kmax}
 \end{figure}

\textit{Results.}---Our main results are shown in \cref{fig:alpha_1D-constraint_FBI_vs_P+B_SNGz05_2kmax}, where we compare $\alpha$ posteriors between FBI and SBI $\PplusB$.
All analyses recover the ground-truth $\alpha=1$ within 68\%CL.
Specifically, at $\kmax=0.1\ (0.12)\iMpch$, FBI analyses constrain $\alpha=0.976\pm0.056$
($\alpha=1.013\pm 0.033$), a 5.9\% (3.6\%) constraint on $\alpha$.
This corresponds to a factor of 3.5 (5.2) improvement over the SBI $\PplusB$ constraints, which are $\alpha=1.014\pm0.200$ ($\alpha=0.872\pm0.170$).
An increase in the improvement of field-level constraints over low-order summary statistics with the analysis cutoff scale $\kmax$ is expected, since the information gain is due to the nonlinearities in the forward model, whose significance increases with wavenumber.

Both FBI and SBI $\PplusB$ results show consistent posteriors between the two $\kmax$ values for each inference method. Their results are further consistent with each other within 0.2-$\sigma$ (0.8-$\sigma$). The consistency between the two analyses (at both $\kmax$) stems from their common forward model, \code{LEFTfield}. The level of consistency further underlines the precision of \code{LEFTfield} on these scales.

\begin{figure}[th]
 %  \centering
   \includegraphics[width=0.8\linewidth]{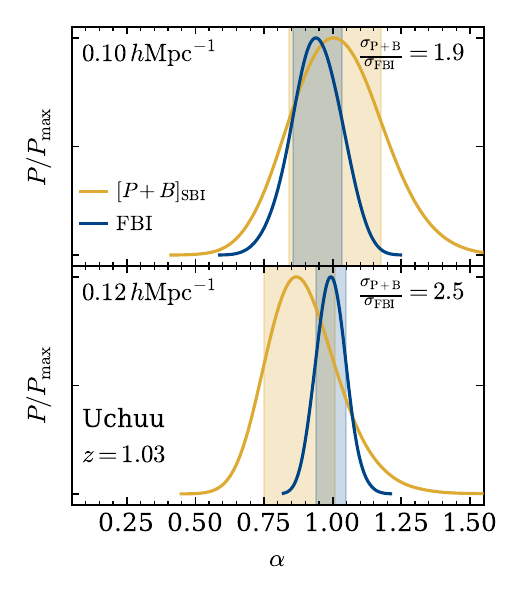}
   \caption{Similar to \cref{fig:alpha_1D-constraint_FBI_vs_P+B_SNGz05_2kmax}, but obtained from the \code{Uchuu} sample (see text), also at $\kmax=[0.10,0.12]\iMpch$.}
   \label{fig:alpha_1D-constraint_FBI_vs_P+B_Uchuuz10_2kmax}
 \end{figure}

To verify whether the above conclusions generalize, we analyze an external sample from the publicly available \code{Uchuu} simulation.
\cref{fig:alpha_1D-constraint_FBI_vs_P+B_Uchuuz10_2kmax} shows that the answer is affirmative: the FBI analysis yields a factor of 1.9 (2.5) improvement over that obtained with SBI $\PplusB$. Specifically, the FBI constraints are $\alpha=0.941\pm0.090$ ($\alpha=0.993\pm0.053$) versus the $\alpha=1.018\pm0.168$ ($\alpha=0.900\pm0.136$) constraints by SBI $\PplusB$, at $\kmax=0.1\ (0.12)\iMpch$, in excellent agreement within 0.4-$\sigma$ (0.6-$\sigma$).
The improvement factors for the Uchuu sample are slightly lower than for the SNG sample. This is likely because displacement contributions to higher $n$-point functions are less significant at higher redshifts (due to a smaller growth factor). As these contributions provide additional information for improving $\sigma_8$ inference at the field level, their reduced importance at higher redshifts could explain the reduced improvements.

In the variant SBI case, which resembles current standard $\PplusB$ analyses, the improvement factors between FBI and SBI $\PplusB$ constraints on $\alpha$ at $\kmax=0.1\ (0.12)\iMpch$ are 3.5 (5.2) for the SNG halo sample and 2.3 (3.5) for the Uchuu halo sample.

\textit{Summary and discussion.}---In this Letter, we have presented the first $\sigma_8$ constraints from field-level inference on fully nonlinear biased tracers, specifically N-body halos in their comoving rest frame. Our constraints are based on the validity of the EFTofLSS on quasilinear scales, and rigorously marginalize over fully nonlinear scales.
While we have not incorporated RSD due to tracer peculiar velocities, which are usually employed to constrain the combination $f\sigma_8$, our analysis remains a useful demonstration: First, a separate $\sigma_8$ constraint, combined with $f\sigma_8$ as inferred from \emph{linear} RSD, allows for a direct inference of the growth rate $f$, which is a sensitive probe of dark energy and gravity \cite{Linder:2005,Linder_Cahn:2007,Linder:2009,Wen:2023bcj}, much more so than the combination $f\sigma_8$.
Second, if a given galaxy sample is affected by line-of-sight-dependent selection bias, then the leading RSD contribution is perfectly degenerate with the leading selection bias term \cite{2009MNRAS.399.1074H,2011ApJ...726...38Z}, i.e. in the same way as $b_1$ is degenerate with $\sigma_8$ in the analysis presented here; however, there are analogous higher-order protected RSD contributions which break the selection bias-growth rate degeneracy \cite{Agarwal:2020lov}.
Hence, even when moving to a redshift-space analysis, being able to break the $b_1-\sigma_8$ degeneracy remains extremely helpful for $\sigma_8$ and $f$ constraints.

We compare our field-level results with a simulation-based inference based on summary statistics, namely the power spectrum and bispectrum. Using the same field-level forward model in both analyses, we demonstrate that the field-level approach significantly outperforms the summary statistics [\cref{fig:alpha_1D-constraint_FBI_vs_P+B_SNGz05_2kmax,fig:alpha_1D-constraint_FBI_vs_P+B_Uchuuz10_2kmax}].
While previous works, e.g. \cite{Schmidt:2018bkr,Cabass:2023nyo}, have shown a correspondence of field-level inference with power spectrum and bispectrum, this only holds when expanding the field-level likelihood at second order in perturbation theory. Here, our forward model includes third-order bias, and is thus expected to incorporate information from higher $n$-point functions. Indeed, our results show that, even on quasilinear scales, there is significant cosmological information beyond the power spectrum and bispectrum.

In future work, we will explore whether additional low-order summaries could extract this information, such as the trispectrum (4-point function).

While we have focused on dark-matter halos here, we demonstrate in \cite{Beyond2pt:preprint} that this conclusion holds for simulated galaxies as well, as expected from EFT principles.
Looking forward to FBI on observed data, \cite{Cabass:2020jqo,Stadler:2023hea} provided the theoretical framework for and demonstrated a successful implementation of RSD into the \code{LEFTfield} forward model, respectively. This should enable a field-level analysis in redshift space and open up the access to additional cosmological information, as argued above.

We stress that we have not attempted to push our analysis to even smaller scales, instead aiming for converged posteriors at conservative scale cuts of $\kmax \leq 0.12\iMpch$ \footnote{The cost of our forward model scales directly as $\sim \Lambda^3\ln\Lambda$.}.  Already in this case, our results indicate that field-level inference enables robust constraints on the growth of structure, independent of the growth rate $f$, at the few-percent level even within a modest volume of $8\, (h^{-1}\mathrm{Gpc})^3$.
This should allow for correspondingly improved constraints on cosmological parameters, in the standard $\Lambda$CDM as well as extended models, using the upcoming DESI \cite{DESI:2016fyo,Goldstein:2022okd}, Euclid \cite{Euclid:2019clj,Euclid:2023bgs} and PFS \cite{PFS:whitepaper2014} data.

\medskip

\begin{acknowledgments}
\textit{Acknowledgments.}---We thank Yosuke Kobayashi, Andres N. Salcedo and Elisabeth Krause for organizing and leading the Beyond-2pt data challenge \github{https://github.com/ANSalcedo/Beyond2ptMock}, which provided vital motivation to finish this work.
This study greatly benefitted from discussions during the Beyond-2pt workshop, organized at \href{https://www.aspenphys.org/}{Aspen Center for Physics} in the summer of 2022.
MN and FS thank the ACP for hospitality.
The workshop and ACP at the time were supported by the National Science Foundation grant PHY-1607611.
We extend our thanks to Dragan Huterer, Mikhail M. Ivanov, Chirag Modi, Julia Stadler and Masahiro Takada for useful discussions during the preparation of this manuscript.
We further thank the two referees for their valuable feedback on our manuscript.
MN and FS acknowledge discussions and collaborations with \href{https://www.aquila-consortium.org/}{Aquila consortium} members on previous works that eventually led to these results.
MN acknowledges the \href{https://www.leinweberfoundation.org/}{Leinweber Foundation} and their support for the \href{https://lsa.umich.edu/lctp}{Leinweber Center for Theoretical Physics}.
Our analyses are performed on the \href{https://docs.mpcdf.mpg.de/doc/computing/clusters/systems/Astrophysics/MPA-FREYA.html}{\code{FREYA}} and \href{https://docs.mpcdf.mpg.de/faq/hpc_systems.html#raven}{\code{RAVEN}} clusters, maintained by the \href{https://www.mpcdf.mpg.de/}{Max Planck Computing \& Data Facility}, as well as the \code{ADA} cluster hosted and maintained by MPA.
We acknowledge the non-negligible \href{https://www.mpcdf.mpg.de/about/co2-footprint}{carbon footprint} of computational research and associated environmental impacts. On an Intel® Xeon® Platinum 8360Y processor, around the Garching, Munich metropolitan area, each \code{LEFTfield} MCMC chain in the FBI analysis averages a footprint of $\sim7\,$mg CO$_{2}$ per MCMC sample.
Our post-processing pipeline is empowered by \href{https://root.cern.ch}{\code{ROOT}}, \href{https://root.cern.ch/manual/python/}{\code{pyROOT}}, \href{https://getdist.readthedocs.io/en/latest/}{\code{GetDist}}, \href{https://numpy.org/}{\code{numpy}}, \href{https://www.dask.org/}{\code{dask}}, and \href{https://matplotlib.org/}{\code{matplotlib}}.
Our collaborative experience is greatly enhanced by the collaborative editor \href{https://hedgedoc.org/}{HedgeDoc}, hosted by the \href{https://gwdg.de/en/services/e-mail-collaboration/gwdg-pad/}{GWDG Pad} service.
Our color choice is inspired by \href{https://personal.sron.nl/~pault/}{Paul Tol's notes} on color blindness and color-blind friendly color schemes.
\end{acknowledgments}

\bibliographystyle{apsrev4-2}
\bibliography{references}

\newpage
\pagebreak

\appendix
\onecolumngrid
% Begin of SM

%TC:ignore
\widetext
\begin{center}
\textbf{\Large Supplementary Material}
\end{center}

%%%%%%%%%%%%%%%%%%%%%%%%%%%%%%%%%%%%%%%%%%%%%%%%%%%%%%%%%%%%%%%%%%%%%%%%%%%%%
\section{Forward model}

The \code{LEFTfield} forward models were detailed in \cite{Schmidt:2020ovm}, specifically their Sec.~2.
Here, we describe the specific procedure we follow to construct the Eulerian density field $\d$ and bias fields $O$ out of a particular realization of the linear density field $\dlin$:
\begin{enumerate}
\item Apply a cubic sharp-$k$ filter with cutoff $\Lambda$ to the linear density field $\dlin \to \dlin_\Lambda$ in Fourier space.
Nonlinear gravitational evolution induces mode coupling, hence contributions from arbitrarily small-scale modes. This cutoff is therefore required for either the Gaussian initial conditions or linear density field \cite{Schmidt:2020viy,Rubira:2023vzw}.
\item Construct the Lagrangian displacement field $\mathbf{s}(\vk)=\sum_n\mathbf{s}^{(n)}(\vk)$, where $n$ is the order of Lagrangian Perturbation Theory (LPT); in this work, we adopt $n=2$, i.e. 2LPT.
\item Evaluate the displacement field and displace mass particles from Lagrangian to Eulerian positions accordingly in real space.
\item Assign the displaced mass particles to the Eulerian grid of size $\NgridEul$ via a non-uniform-to-uniform fast Fourier transform (NUFFT, see below).
\item Apply another cubic sharp-$k$ filter with cutoff $\Lambda' = 1.2\kmax$.
\item Construct the bias fields $O$ out of the filtered Eulerian density grid $\d_{\Lambda'}$. We choose $\Lambda=1.2\Lambda'=1.44\kmax$ throughout.
\end{enumerate}

A new feature of \code{LEFTfield} in this work is the introduction of the NUFFT algorithm for density assignment in step 4 above. NUFFT implements the nonuniform-to-uniform discrete Fourier transform $f(\vec x) \to \tilde f(\vec k)$ by assigning pseudoparticles at positions $\vec x_i$ with weights $f(\vec x_i)$ to a supersampled grid $(\NgridNUFFT)^3$ (typically $\NgridNUFFT=1.2-2\NgridEul$) using a specific assignment kernel with compact support (roughly 4-16 grid cells). The algorithm proceeds to perform an FFT on the supersampled grid, deconvolves the assignment kernel, and resizes the grid in Fourier space to the target uniform discrete Fourier transform $\tilde f(\vec k)$. This method is approximate, but accuracy close to machine precision can be obtained for very reasonable computational effort \cite{NUFFT}.

%%%%%%%%%%%%%%%%%%%%%%%%%%%%%%%%%%
\subsection{Galaxy bias expansion}

The EFT galaxy bias expansion expands the local galaxy density field $\dg$ in a set of galaxy bias operators, i.e. fields, constructed recursively out of the gravitational potential $\Phi$. The latter relates to the local matter density field $\d$ through the Poisson equation
\be
\nabla^2\Phi(\vx,\tau) = (aH)^2\Om\d(\vx,\tau),
\label{eq:Poisson_eq}
\ee
in which the scale factor $a$ is a function of the conformal time $\tau$.

At each $n$-th order, the set includes local gravitational operators that can be classified into two categories: (1) operators that involve two spatial derivatives for each instance of $\Phi$ and combinations thereof, plus (2) higher-derivative operator(s) that involve more than two spatial derivatives acting on a single instance of $\Phi$ \cite{Desjacques:2016bnm}.

Let us first consider category (1). Following \cite{Mirbabayi:2014zca,Desjacques:2016bnm} and Einstein notation, we define
\be
\Pi^{[1]}_{ij}(\vx,\tau) = \frac{2}{3\Om aH^2} \partial_{x_i}\partial_{x_j}\Phi(\vx,\tau) = K_{ij}(\vx,\tau)+\frac{1}{3}\d^K_{ij}\d(\vx,\tau),
\label{eq:Pi_def}
\ee
where $\mathbf{K}\equiv\Del\d$ denotes the local tidal field tensor and $\d^K$ denotes the Kronecker delta.
The superscript $[1]$ indicates that the lowest-order contribution to this operator is at $n=1$ order in perturbation theory.

It follows that $\mathbf{\Pi}^{[n]}$ is given by the recursive relation:
\be
\Pi^{[n]}_{ij} = \frac{1}{(n-1)!} \left[(aH f)^{-1}\convD \Pi^{[n-1]}_{ij} - (n-1) \Pi^{[n-1]}_{ij}\right],
\label{eq:Pi_recursive_def}
\ee
where $\convD$ is the convective derivative with respect to $\tau$.

Up to third order in the Eulerian basis, the complete set of operators in category (1) is given by
\ba
&\text{1st-order} \, &&\tr\left[\mathbf{\Pi}^{[1]}\right]\\
&\text{2nd-order} \, &&\tr\left[\mathbf{\Pi}^{[1]}\,\mathbf{\Pi}^{[1]}\right],
\tr\left[\mathbf{\Pi}^{[1]}\right]\,\tr\left[\mathbf{\Pi}^{[1]}\right],\\
&\text{3rd-order} \, &&\tr\left[\mathbf{\Pi}^{[1]}\,\mathbf{\Pi}^{[1]}\,\mathbf{\Pi}^{[1]}\right],
\tr\left[\mathbf{\Pi}^{[1]}\,\mathbf{\Pi}^{[1]}\right]\tr\left[\mathbf{\Pi}^{[1]}\right],
\tr\left[\mathbf{\Pi}^{[1]}\right]\,\tr\left[\mathbf{\Pi}^{[1]}\right]\,\tr\left[\mathbf{\Pi}^{[1]}\right],
\tr\left[\mathbf{\Pi}^{[1]}\mathbf{\Pi}^{[2]}\right].
\ea
The above list, albeit complete (at third order), is only one possible linear combination of Eulerian cubic bias operators $O$ in category (1).
In fact, there exists different equivalent linear combinations and conventions in the literature, e.g. \cite{McDonald:2009dh,Assassi:2014fva}. We refer readers to App.~C.2 of \cite{Desjacques:2016bnm} for further details and discussions.

The full set of EFT bias operators $O$ up to third order, following the convention in \cite{Mirbabayi:2014zca,Desjacques:2016bnm}, therefore is
\be
O\in\left[\d,\d^2,K^2,\d^3,K^3,\d K^2,\Otd,\nabla^2\d\right],
\label{eq:Eulerian_third_bias_O}
\ee
where we include one operator from category (2), namely the leading-order higher-derivative bias operator $\nabla^2\d$.
At third order, a nonlocal operator from category (1) appears in \cref{eq:Eulerian_third_bias_O}, namely
\be
\Otd\equiv\frac{8}{21}K^{(1)}_{ij}\mathcal{D}^{ij}\left[\left(\dlin\right)^2-\frac{3}{2}\left(K^{(1)}_{ij}\right)^2\right].
\label{eq:Otd_def}
\ee
\cref{eq:Eulerian_third_bias_O} explicitly provides the EFT bias operators considered in the modeling of the FBI analysis and SBI $\PplusB$ Fisher forecast.

For both inference methods (FBI and SBI $\PplusB$), we use \code{LEFTfield} to construct the Eulerian galaxy bias fields $O$ in \cref{eq:Eulerian_third_bias_O} out of the filtered Eulerian matter density field $\d_{\Lambda'}$ at the tracer redshift $z$, following the procedure outlined in the previous section of Supplementary Material.

%%%%%%%%%%%%%%%%%%%%%%%%%%%%%%%%%%
\subsection{Marginalization over galaxy bias}
\label{subsec:FBI_marg_likelihood}

\cref{eq:FBI_likelihood} can be rewritten as
\ba
&\L_\fbi^\expl\left(\d^\obs_{g}  \Big| \shat, \alpha, \{b_O\}, \{\sigmaEps\}\right) = \mathbb{N}
\exp\left[-\frac{1}{2} \sum_{\vk>0}^{\kmax} \ln \sigmaEps^2(k)\right] \times \exp\left\{
- \frac{1}{2} C
+ \sum_{O} b_O B_O
- \frac{1}{2} \sum_{O,O'} b_O b_{O'} A_{OO'} \right\} \, ,
\numberthis
\label{eq:FBI_premarglikelihood}
\ea
where $\mathbb{N}$ is a parameter-independent normalization constant and following \cite{Elsner:2019rql}, we introduce a scalar $C$, a vector $B_O$ and a matrix $A_{OO'}$:
\ba
C &= \sum_{\vk>0}^{\kmax} \frac{1}{\sigmaEps^2(k)} \left| \d^\obs_{g} - \sum_{O_\unmarg}b_{O_\unmarg}O_\unmarg(\vk) \right|^2\\
B_O &= \sum_{\vk>0}^{\kmax} \frac{\left[\d^\obs_{g}(\vk) -\sum_{O_\unmarg}b_{O_\unmarg}O_\unmarg(\vk)\right] O^*(\vk)}{\sigmaEps^2(k)} \\
A_{OO'} &= \sum_{\vk>0}^{\kmax} \frac{O(\vk) O'^*(\vk)}{\sigmaEps^2(k)}. \numberthis
\label{eq:ABC}
\ea
\cref{eq:FBI_premarglikelihood} thereby separates the $n$ galaxy bias coefficients $\{b_O\}$  to be marginalized over and the rest $\{b_O\}_\unmarg$ to be left unmarginalized.

For Gaussian priors assumed in \cref{eq:FBI_priors}, integrating out $\{b_O\}$ in \cref{eq:FBI_premarglikelihood} (using Gaussian integrals) yields
\ba
&\L_\fbi^\expl\left(\d^\obs_{g}  \Big| \shat, \alpha, \{b_O\}_\unmarg, \{\sigmaEps\}\right) = \left( \prod_O \int db_O\right) P \left(\d^\obs_{g} \Big| \shat, \{b_O\} \right) \\
&= \mathbb{N} (2\pi)^{n/2} \Big| A_{O O'} \Big|^{-1/2}
\exp \left[ -\frac{1}{2} \sum_{\vk>0}^{\kmax} \ln \sigmaEps^2(k) \right] \\
& \times \exp \left\{ -\frac{1}{2} C(\{b_O\}) + \frac{1}{2}
\sum_{O,O'} B_O(\{b_O\}) (A^{-1})_{O O'} B_{O'}(\{b_O\}) \right\} \, .
\numberthis
\label{eq:FBI_marglikelihood}
\ea
In the FBI analysis, we analytically marginalize the entire set of bias coefficients $\{b_O\}$.
Analytical marginalization improves the MCMC sampling efficiency by reducing the dimensions of parameter space to explore.
A validation of \cref{eq:FBI_marglikelihood} on mock data can be found in App.~F of \cite{Kostic:2022vok}.

\begin{figure}[t]
   \centering
 \includegraphics[width=0.7\linewidth]{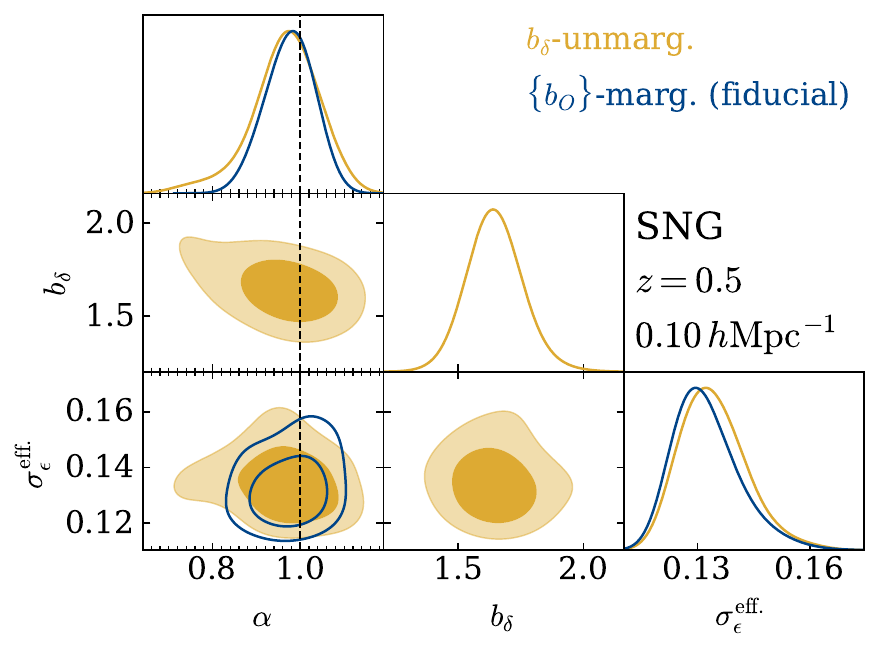}
   \caption{Posterior consistency between the (fiducial) $\{b_O\}$-marginalized likelihood (blue, open) and the $b_\d$-unmarginalized likelihood (yellow, filled). Panels show the 1D and 2D constraints on the amplitude rescaling parameters $\alpha=\sigma_8/\sigma_{8,\fid}$, the linear galaxy bias coefficient $b_\d$, and the galaxy stochasticity parameters $[\sigma_{\eps,0},\sigma_{\eps,k^2}]$. Contours indicate the 68\% and 95\% confidence intervals.}
   \label{fig:FBI_corner_z05_kmax010}
 \end{figure}

In this work, we validate the analytical marginalization by re-analyzing the SNG sample at $\kmax=0.10\iMpch$ using a likelihood that marginalizes over all galaxy bias coefficients in \cref{eq:Eulerian_third_bias_O} except the linear bias coefficient $b_\d$. In \cref{fig:FBI_corner_z05_kmax010}, we compare the posteriors obtained with this ``$b_\d$-unmarg.'' likelihood with those obtained in our fiducial FBI analysis using the ``$\{b_O\}$-marg.'' likelihood [\cref{eq:FBI_marglikelihood}].  All 1D and 2D marginal posteriors are consistent with their corresponding counterparts. The $b_\d$-unmarginalized posterior, as projected down to the $b_\d-\alpha$ plane, showcases the ability of FBI to break the $b_\d-\sigma_8$ degeneracy by extracting information from nonlinear clustering: both $b_\d$ and $\alpha$ are well constrained.

We note however that the ``$b_\d$-unmarg.'' analysis assumes a prior consisting of the product of the fiducial Gaussian prior (see next section) and a uniform prior, i.e. $\P(b_\delta|{\rm unmarg}) = \N(1.0,5.0)\, \U(0.,7.)$, and is thus slightly mismatched with the marginalized chains. However, given the tight constraint obtained on $b_\delta$, we do not expect this small mismatch to be quantitatively relevant. However, the ``$b_\d$-unmarg.'' chains have only reached roughly 21 effective samples, as compared to the substantially larger sample size of the fiducial chains [\cref{tab:EFT-fieldlevel_neff}]. We therefore only focus on the overall consistency between the ``$b_\d$-unmarg.'' and the fiducial ``$\{b_O\}$-marg.'' posteriors.
%

%%%%%%%%%%%%%%%%%%%%%%%%%%%%%%%%%%
\subsection{Priors}
\label{subsec:Priors}

\paragraph{FBI priors.}
In the FBI analysis, we assume the following priors on the amplitude rescaling parameter $\alpha$, bias coefficients $\{b_O\}$ in \cref{eq:Eulerian_third_bias_O}, and the noise parameters $\{\sigmaEps\}$:
\begin{gather*}
\P(\alpha) = \U(0.5,1.5), \\
\P(b_\delta) = \N(1.0,5.0), \\
\P(b_{\delta^2}) = \P(b_{K^2}) = \P(b_{\delta^3}) = \P(b_{\delta K^2}) = \P(b_{K^3}) = \P(b_{\Otd}) = \N(0.0,1.0), \\
\P(b_{\nabla^2\delta}) = \N(0.0,5.0), \\
\P(\sigma_{\eps,0}) = \U(0.8\sigma_{\eps,\mathrm{Poisson}},100.), \quad \P(\sigma_{\eps,k^2}) = \U(-10.0,100.0),\numberthis\label{eq:FBI_priors}
\end{gather*}
where $\U(a,b)$ represents a uniform distribution between the lower bound $a$ and upper bound $b$, while $\N(\mu,\sigma)$ represents a normal distribution with mean $\mu$ and standard deviation $\sigma$.

$\sigma_{\eps,\mathrm{Poisson}}$ in \cref{eq:FBI_priors} is the expected (Gaussian) RMS noise amplitude based on the Poisson shot noise expectation given the comoving number density of the tracers considered. We find it necessary to impose an informative lower limit on the noise level, as the FBI MCMC chains tend to get stuck in a region of unphysically low $\sigmaEps$. The causes of this drift are still under investigation. The lower limit $0.8\sigma_{\eps,\mathrm{Poisson}}$ we impose in this work is significantly \emph{below} what previous studies of halo stochasticity have found \cite{2010PhRvD..82d3515H,Schmittfull:2018yuk}.

We note that our priors on $\{b_O\}$ are consistent with other EFT cosmology analyses, e.g. \cite{Ivanov:2021kcd,Philcox:2021kcw,Beyond2pt:preprint}.
Further, in \cite{Beyond2pt:preprint}, we have explicitly verified that uniform flat priors (see their Eqs.~(20-21)), Gaussian priors (our Eq.~(A9)), or anything in between, on $\{b_O\}$ do not affect the $\sigma_8$ constraint.

\paragraph{SBI $\PplusB$ priors.}
The SBC tests for asserting the uncertainties of the estimated SBI posterior are only feasible when the posterior is amortized, i.e., not restricted to a single observation. However, performing an amortized NPE by drawing samples from the considerably wide FBI prior would require a huge amount of simulations for convergence.

We therefore follow the same strategy described in \cite{Tucci:2023bag}: (1) We initiate the training sequence using the same priors as in the FBI analysis and a sequential NPE (SNPE) is performed over multiple rounds, where the posterior of each round is used as proposal for the next one. This allows us to exclude the prior regions where the posterior has negligible support. (2) The training data for the final NPE posterior is then sampled from a wider Gaussian than the posterior obtained in the last SNPE round. This allows for faster SBC analysis and for convergence with fewer simulations, but still guarantees that both SBI and FBI analysis are consistent and that none of the parameters are prior dominated.

Another subtlety employed in SBI is that the bias parameters $b_{O^{(n)}}$, where $n$ stands for the order of the bias operators as listed in \cref{eq:Eulerian_third_bias_O}, are correspondingly scaled to $\alpha^nb_{O^{(n)}}$ when doing inference. This avoids prior volume effects by reducing degeneracies from the SBI data vector. Additionally, the higher-derivative biases $b_{\nabla^2\delta}$ and $b_{\nabla^2\varepsilon}$ are made dimensionless after a scaling by $R_{\ast}^{-2}$, where $R_{\ast}$ is the characteristic scale of halo formation (in this work, we assume $R_{\ast}=5h^{-1}\mathrm{Mpc}$).

%%%%%%%%%%%%%%%%%%%%%%%%%%%%%%%%%%
\subsection{Eulerian and Lagrangian bias expansions}
\label{subsec:Eulerian-Lagrangian}

In both FBI and SBI $\PplusB$ fiducial analyses, we adopt the Eulerian basis for the EFT galaxy bias expansion, as described in Sec.~II of the Supplementary Material. Another valid choice of basis for the EFT bias fields $O$ and bias expansion is the Lagrangian space \footnote{See Sec.~2.5.2-2.5.3 in \cite{Desjacques:2016bnm}.}. In principle, we can construct $O$ at the Lagrangian coordinates $\vq=\vx(\tau=0)$ and advect them to the final Eulerian coordinates $\vx(\vq,\tau)$, following the procedure in Sec.~3 of \cite{Schmidt:2020ovm}.
As the displacement field contains only perturbative modes and is protected by the equivalence principle of corrections from bias terms, UV effects in displacement operations are not of concern. Therefore, a cut in the evolved density field, as listed by step 5 in App.~A1, is unnecessary in the Lagrangian-bias forward model.

Both bias expansions are equivalent at fixed order in perturbation theory but differ in higher-order terms. Thus, a comparison of both cases provides an estimate of the importance of higher-order terms in the bias expansion.

\begin{figure}[t]
   \centering
 \includegraphics[width=0.7\linewidth]{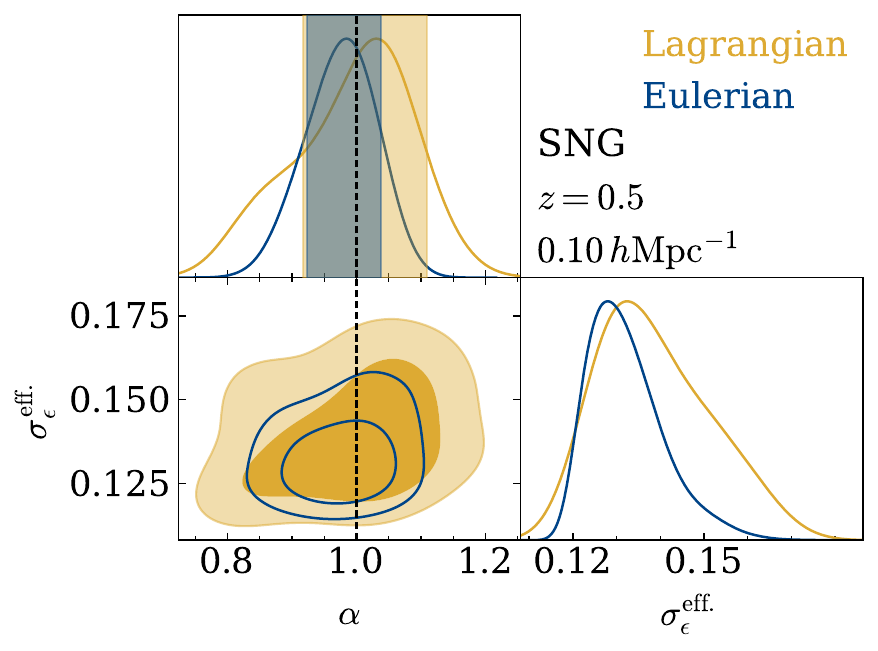}
   \caption{Posterior consistency between the (fiducial) Eulerian bias and the Lagrangian bias. Constraints on the amplitude rescaling parameter $\alpha=\sigma_8/\sigma_{8,\fid}$ and the effective noise amplitude $\sigmaEps^{\eff}=\sigma_{\eps,0}[1+\sigma_{\eps,k^2}\kmax^2]$, obtained with the fiducial Eulerian bias model (blue, open) and with the alternative Lagrangian bias model (yellow, filled). Contours indicate the 68\% and 95\% confidence intervals of the corresponding 2D marginal posteriors. Vertical bands indicate the 68\% limits of the corresponding 1D marginal posteriors.}
   \label{fig:alpha-sigmaEps_2D-constraint_Eulerian_vs_Lagrangian_kmax010}
 \end{figure}

In \cref{fig:alpha-sigmaEps_2D-constraint_Eulerian_vs_Lagrangian_kmax010}, we show the constraints obtained with each bias model, for the fiducial halo sample at $z=0.5$ and the analysis cutoff scale $\kmax=0.10\iMpch$. Posteriors from the two analyses are consistent with one another, taking into account the fact that the Lagrangian chains only consist of roughly 13 effective samples in total (and hence moments of its estimated posteriors are not expected to be accurate).

%%%%%%%%%%%%%%%%%%%%%%%%%%%%%%%%%%%%%%%%%%%%%%%%%%%%%%%%%%%%%%%%%%%%%%%%%%%%%
\section{Posterior sampling and validation}

\subsection{FBI---MCMC chain initialization}
\label{subsec:MCMC_initialization}

Regardless of the initialization, once converged, MCMC chains should sample the same underlying distribution.
MCMC convergence is particularly relevant in the case of high-dimensional input data and sampling space, such as that of FBI.

In this work, we explore two strategies to initialize a MCMC chain. Specifically, for each FBI analysis, we initialize one MCMC chain from the true initial conditions (true-phase initialization, TPI) $\shat_{\mathrm{true}}$, and multiple chains at different random initial conditions (random-phase initialization, RPI) $\shat$.
Monitoring and comparing behaviors between TPI and RPI chains help us verify that the chains have converged.

\begin{figure}[t]
   \centering
 \includegraphics[width=\linewidth]{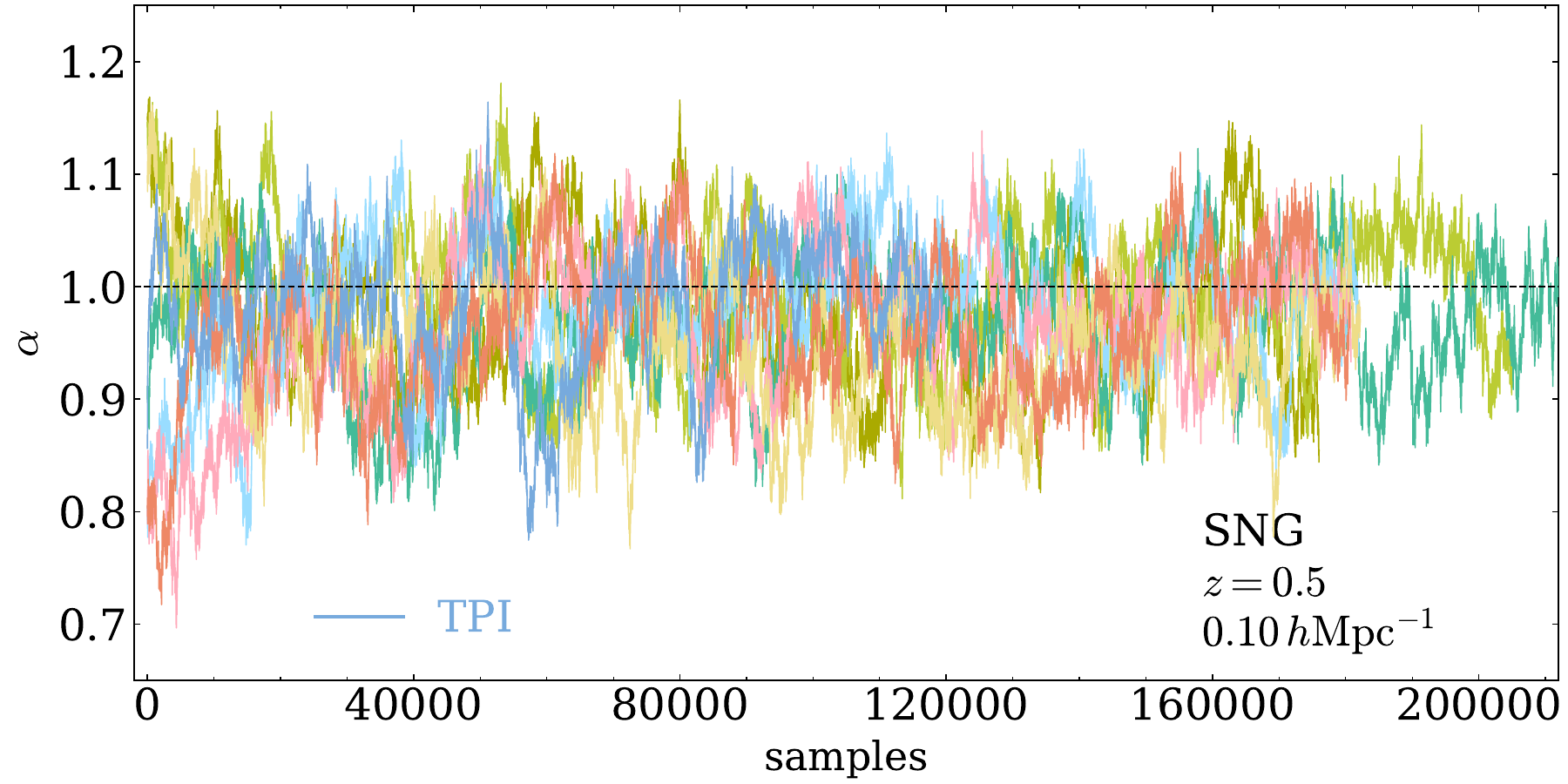}
   \caption{Trace plot of the amplitude rescaling parameter $\alpha=\sigma_8/\sigma_{8,\fid}$ for TPI and RPI chains in the FBI analysis. Different colors indicate chains starting at distinct initial points in parameter space.}
   \label{fig:alpha_trace_TPI+RPIs}
 \end{figure}

In \cref{fig:alpha_trace_TPI+RPIs}, we show a trace plot of the parameter $\alpha$ in the FBI chains, both TPI (light blue) and RPI. After the warm-up phase, they are virtually indistinguishable, all sampling the underlying posterior.

%%%%%%%%%%%%%%%%%%%%%%%%%%%%%%%%%%
\subsection{FBI---Posterior of initial conditions}
\label{subsec:IC_convergence}

\begin{figure}[th!]
\centering
\includegraphics[width=.8\linewidth]{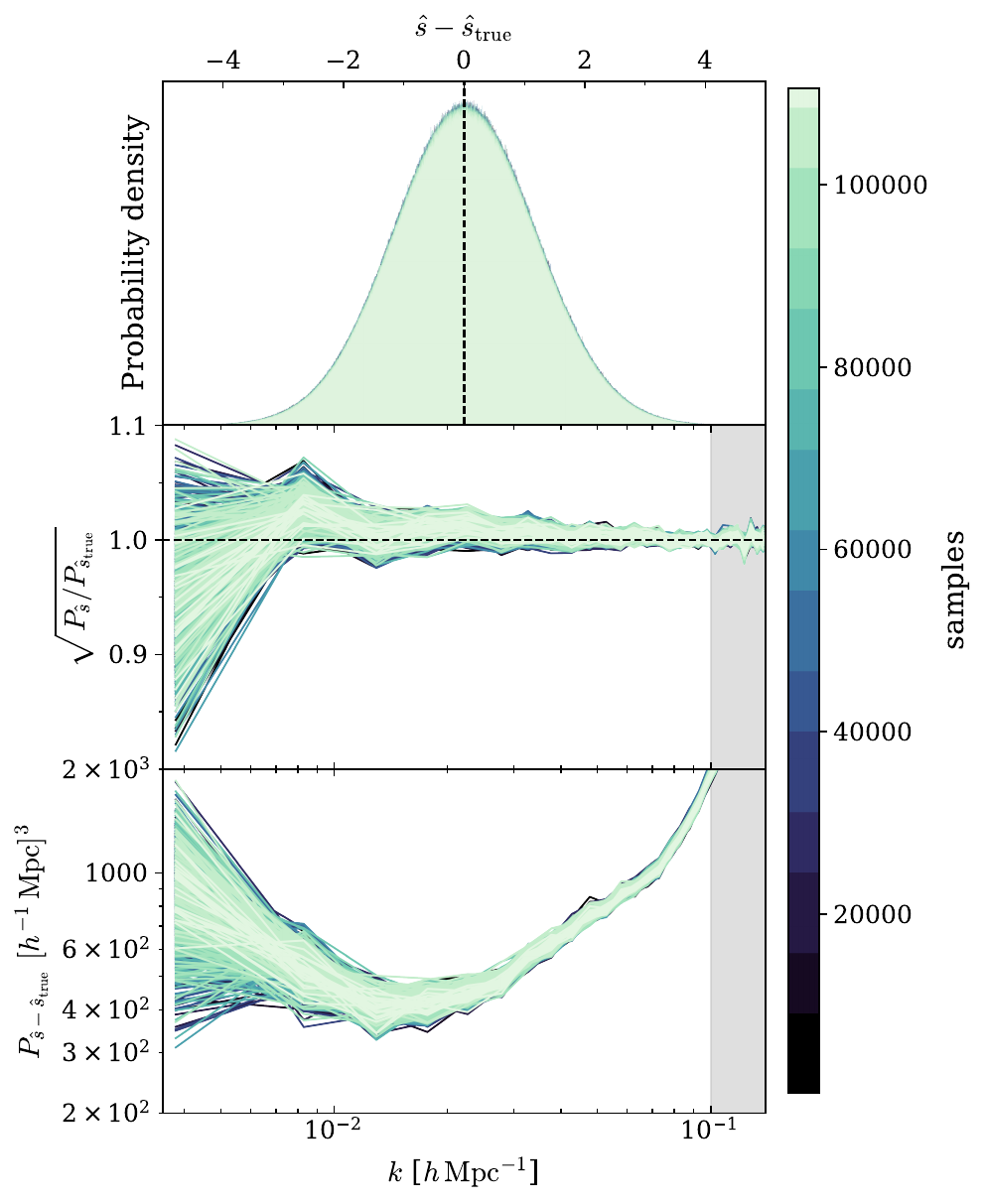}
\caption{Trace plots of different statistics of the posterior initial conditions $\shat$ from our TPI chain in \cref{fig:alpha_trace_TPI+RPIs}. For visibility, in this plot we remove the warm-up phase and further thin the samples by a factor of 100. The grey vertical bands in the lower two panels indicate scales above the cutoff $\kmax=0.1\iMpch$. Top: Distribution of the difference between posterior and true initial conditions, $\shat-\shattrue$. Middle: Square root of the ratios between posterior and true power spectrum of initial conditions $P_{\shat}/P_{\shattrue}$. Bottom: Variance of the residuals between posterior and true initial conditions, $P_{\shat-\shattrue}$.}
\label{fig:shat-dist_Tk_residual-variance_trace}
\end{figure}

Our field-level constraints include the initial conditions $\shat$ of the observed universe volume. \cref{fig:shat-dist_Tk_residual-variance_trace} displays the coverage of the initial condition posterior, in different summary statistics.

Firstly, the histograms in the top panel represent the probability density of $\shat-\shattrue$ where $\shat$ are samples from the posterior, while $\shattrue$ is the true initial conditions. The distribution perfectly follows a normal distribution $\N(0,1)$, indicating that our $\shat$ posterior in FBI correctly recovers the first two moments of the true underlying $\P(\shat|\dg^\obs)$.

Secondly, the middle panel shows square roots of the ratios between posterior $\shat$ power spectra over the $\shattrue$ power spectrum. The quantity---representing a transfer function---is consistent with 1 across all Fourier modes, as expected.

Lastly, in the bottom panel, we show the variance of $\shat-\shattrue$, which goes to zero for zero noise and a perfect forward model. The inferred variance is broadly consistent with the Poisson noise expectation of the data; see also the extended discussion in \cite{Kostic:2022vok}.

%

%%%%%%%%%%%%%%%%%%%%%%%%%%%%%%%%%%
\subsection{FBI---Parameter posterior convergence}
\label{subsec:param_convergence}

\begin{figure}[th!]
\centering
\includegraphics[width=.8\linewidth]{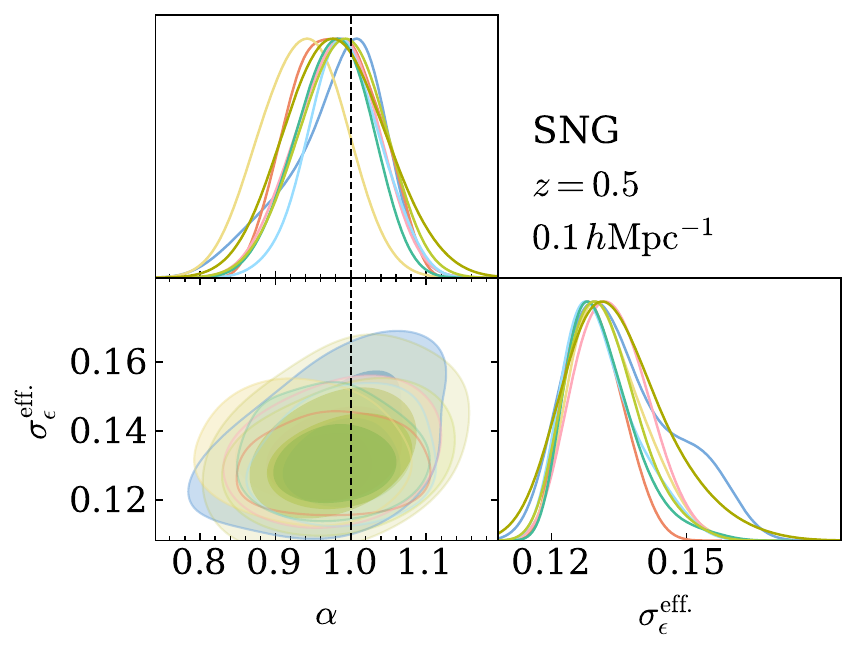}
\caption{Posterior consistency between MCMC chains with different initializations. Each color maps to the chain with the same color in \cref{fig:shat-dist_Tk_residual-variance_trace}. Panels show 1D and 2D marginal posteriors of $[\alpha,\{\sigmaEps^\eff\}]$. Contours indicate the 68\% and 95\% credible intervals.}
\label{fig:alpha_sigmaEps_2D-constraint_FBI_posterior_convergence}
\end{figure}

Given the (extremely) high-dimensional space in FBI, posterior convergence is a particular concern. In \cref{fig:alpha_sigmaEps_2D-constraint_FBI_posterior_convergence}, we show the marginal posteriors of $\alpha$ and $\sigmaEps^\eff=\sigma_{\eps,0}[1+\sigma_{\eps,k^2}\kmax^2]$ from the same set of MCMC chains shown in \cref{fig:alpha_trace_TPI+RPIs}. These chains come from the FBI analysis of the main sample at the analysis cutoff scale $\kmax=0.1\iMpch$. The posterior contours are consistent with each other, indicating that these chains ----individually and jointly---sample the same underlying posterior distribution.

Markov chain transitions and samples are correlated. Therefore, the MCMC sampling error on the mean of $\alpha$ can be estimated by $\sigma_\alpha/\mathrm{ESS}$ where ESS is the effective number of independent samples, i.e. effective sample size of $\alpha$. \cref{tab:EFT-fieldlevel_neff} reports the ESS for each FBI analysis in \cref{fig:alpha_1D-constraint_FBI_vs_P+B_SNGz05_2kmax,fig:alpha_1D-constraint_FBI_vs_P+B-nG_Uchuuz10_2kmax}.

\begin{table}
    \centering
    \begin{tabular}{lcc}
        \hline
        \hline
        Analysis & ESS [samples] \\
        \hline
    \code{SNG}, $\kmax=0.10\iMpch$  & 200 \\
    \code{SNG}, $\kmax=0.12\iMpch$  & 100 \\
    \code{Uchuu}, $\kmax=0.10\iMpch$  & 180 \\
    \code{Uchuu}, $\kmax=0.12\iMpch$  & 60 \\
        \hline   
        \hline
    \end{tabular}
    \caption{Estimates for the effective sample size of $\alpha$ (middle column) and the classical G-R statistics (right column) in the FBI analyses.}
    \label{tab:EFT-fieldlevel_neff}
\end{table}

\subsection{SBI P+B---Posterior diagnostics}\label{subsec:SBIconvergence}

\paragraph{Analysis settings.}
For the scale cuts considered in this work, namely $\kmax=0.1\  (0.12)\iMpch$, the power spectrum bins have dimension $N_{\mathrm{bins}}=15\ (18)$, corresponding to a total of $D=443\ (714)$ dimensions for the SBI data vector. We use a total simulation budget of $\Nsim=5\times10^5$ for all posteriors shown with the except of the SNG halo sample at higher cutoff, which uses $\Nsim=10^6$, as motivated by the convergence tests below. We use $10^5$ samples from the posterior for plotting.

\paragraph{Hyperparameters and training.}
We use the SNPE method of \cite{Greenberg:2019} with 10 atoms for atomic proposals and masked autorregressive flows \cite{Papamakarios:2017} with 10 autoregressive layers, each constructed using two fully-connected $\tanh$ layers with 100 hidden units. We train the models by stochastically minimizing the loss using the Adam optimizer \cite{Kingma:2014} with learning rate of $5\times10^{-4}$ and batch size of 50. 10\% of the samples are used for validation and we stop training if the validation set loss did not improve after 20 consecutive epochs. 

\begin{figure}[t]
   \centering
   \includegraphics[width=\linewidth]{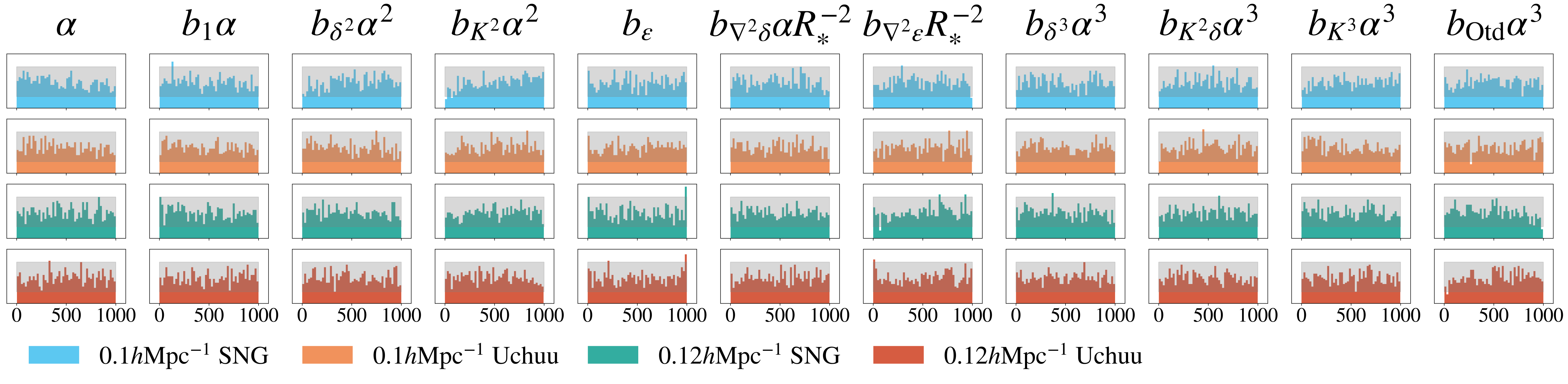}
   \caption{Rank distributions from the SBC tests for the SBI posteriors shown in \cref{fig:alpha_1D-constraint_FBI_vs_P+B_SNGz05_2kmax} and \cref{fig:alpha_1D-constraint_FBI_vs_P+B_Uchuuz10_2kmax}. The grey shaded area indicates the 99\% confidence interval of an uniform distribution. Upper, middle and lower panels correspond to the SNG ($\kmax=0.1\iMpch$), Uchuu ($\kmax=0.1\iMpch$), and SNG ($\kmax=0.12\iMpch$) analyses, respectively.}
    \label{fig:SBI_SBC_Uchuu_SNG_bepsdelta_dpi600}
 \end{figure}

\paragraph{Simulation-based calibration (SBC).}
We use SBC \cite{Talts:2018} to analyze the uncertainties of the final estimated posterior.
A healthy posterior should lead to uniformly distributed rank statistics for all parameters. If the SBI underestimates (overestimates) the true posterior variance for some parameter, one expects a ``U-shaped'' (``$\cap$-shaped'') rank distribution. As shown in \cref{fig:SBI_SBC_Uchuu_SNG_bepsdelta_dpi600}, the ranks are uniformly distributed, indicating that the estimated posteriors passed the calibration test.

\begin{figure}[t]
   \centering
   \includegraphics[width=0.5\linewidth]{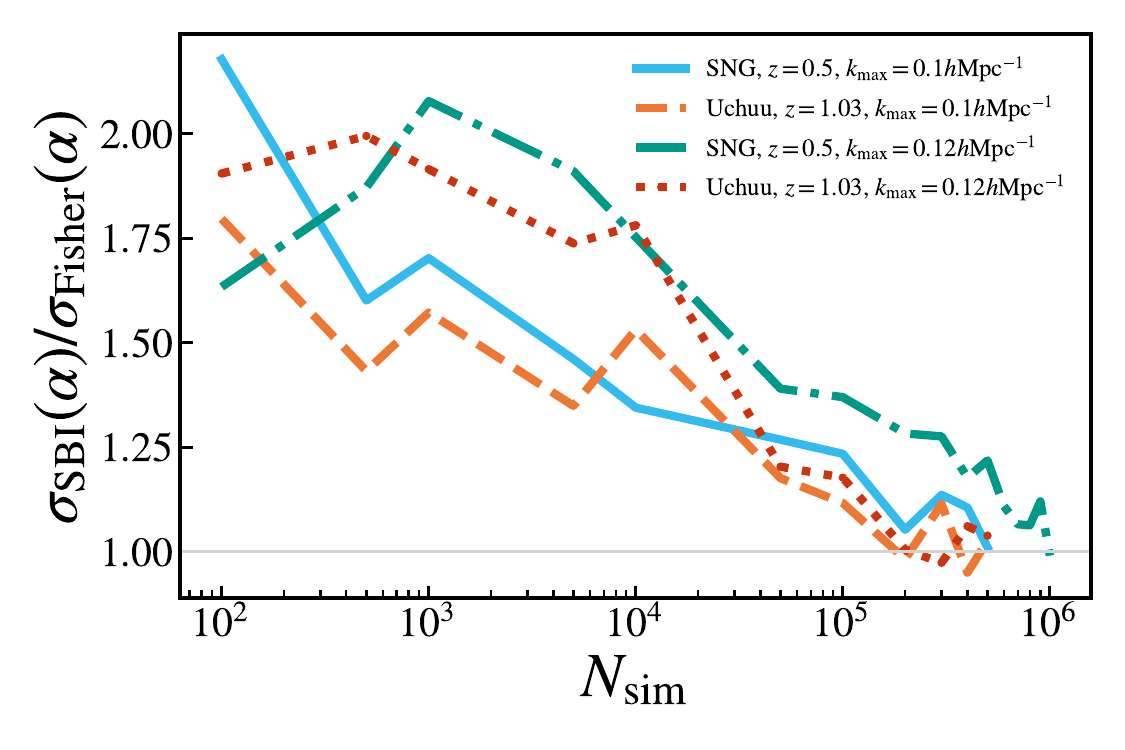}
   \caption{Convergence of the standard deviations of $\alpha$ posteriors in SBI $\PplusB$ analyses with increasing simulation budget $\Nsim$ used for posterior estimation. Each value is normalized by the corresponding $\alpha$ constraint from a Fisher analysis, for each of the cases considered.} 
    \label{fig:convergence_SBI_Rockstar_Uchuu}
 \end{figure}

\paragraph{Convergence tests.}
For SBI, we check for posterior convergence (and uncertainty quantification) by monitoring the estimated uncertainties on $\alpha$ as a function of the simulation budget, $\Nsim$. 
We show the SBI convergence tests in \cref{fig:convergence_SBI_Rockstar_Uchuu}, where the standard deviations of $\alpha$ posteriors are normalized by the same constraints from the corresponding Fisher analyses. We perform this test for all parameters in the SBI $\PplusB$ analyses, but show only the case of $\alpha$ to avoid clutter.

\begin{figure}[t]
   \centering
   \includegraphics[width=0.9\linewidth]{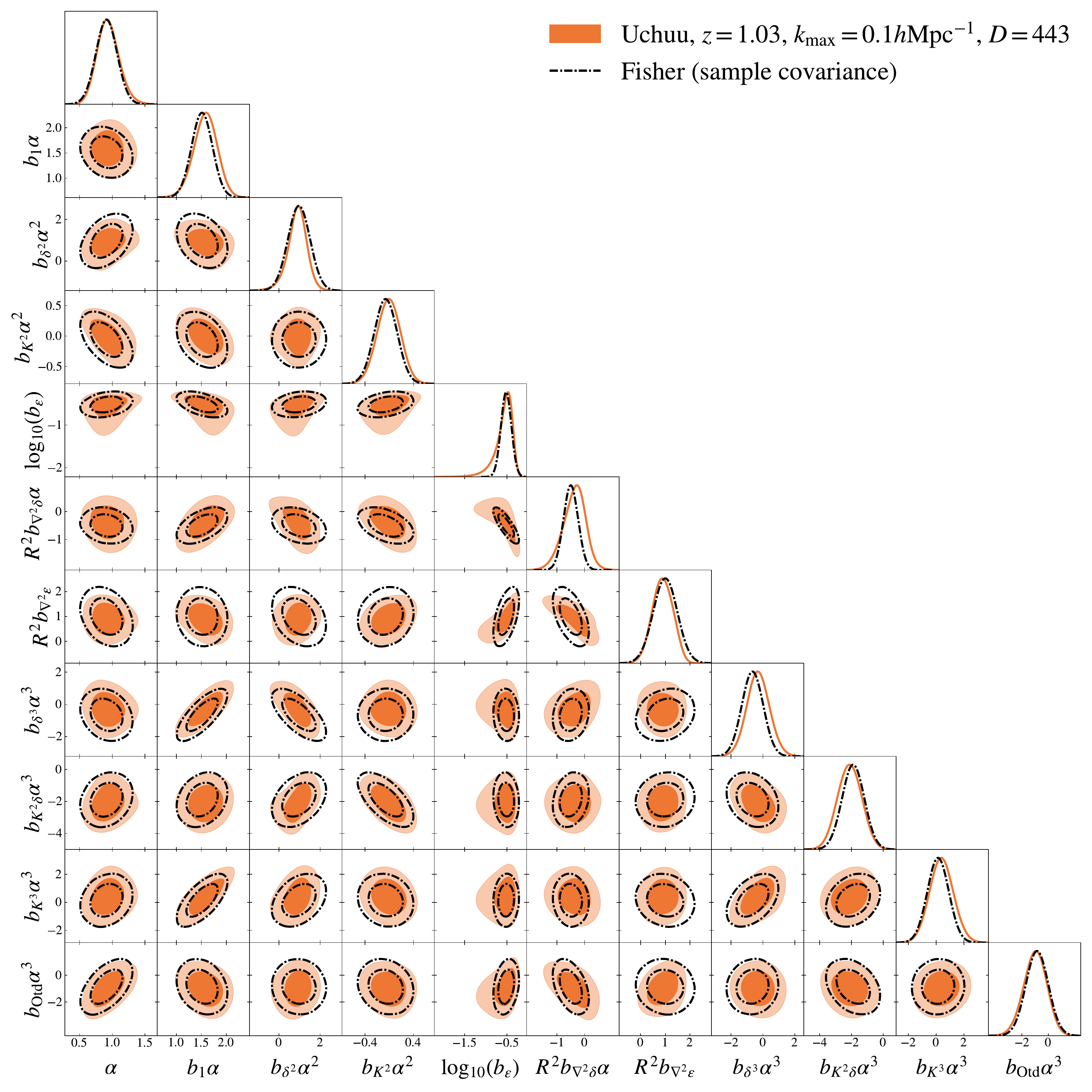}
   \caption{Parameter posteriors obtained in the SBI $\PplusB$ analysis of the Uchuu halo sample at redshift $z=0.5$ with $\kmax=0.1\iMpch$. Contours indicate 68\% and 95\% credible intervals. Dashed lines correspond to the Fisher analysis for this case, using sample covariance for the data vector.
   }
    \label{fig:SBI_2LPTthird_bepsdelta_SNG_z05_L014_posterior_step10_sample}
 \end{figure}

\paragraph{Fisher analysis.}
The $1\sigma$ error on the parameter $\theta_\alpha$ is estimated by $\sqrt{(F^{-1})_{\alpha\alpha}}$, where $F$ is the Fisher information matrix, evaluated at the SBI posterior mode, $\theta_{\textsc{map}}$. Following \cite{Tucci:2023bag}, we compute $F$ by taking numerical derivatives with respect to the model parameters of the mean of the data vector obtained by averaging over 1000 \code{LEFTfield} data realizations. We estimate the sample covariance from $10^5$ simulations evaluated at $\theta_{\textsc{map}}$. For the Fisher analysis when including all third-order bias parameters, we use the $\theta_{\textsc{map}}$ obtained by SBI from the restricted case (i.e., with only $b_{\mathrm{Otd}}$), and set the fiducial values of the other third-order bias parameters to zero. We use one standard deviation of the SBI posterior as step size for the derivatives of the parameters corresponding to the restricted case, and unity as step size for the other third-order bias parameters, motivated by the SBI priors. We also sum the Gaussian FBI priors to the Fisher matrix. We test convergence of the Fisher errors with respect to step size, fiducial point, number of simulations for covariance estimation and for the mean in the derivatives (not shown). We show in \cref{fig:SBI_2LPTthird_bepsdelta_SNG_z05_L014_posterior_step10_sample} the posterior obtained by SBI for one of the cases considered in this work and a comparison to a Fisher analysis using sample covariance. $\log(b_{\epsilon})$ (with basis 10) is shown instead of $b_{\epsilon}$ since its prior is bounded at zero, while Fisher assumes Gaussianity of the parameter posteriors.


\subsection{SBI P+B---Non-Gaussian stochasticity}\label{subsec:SBInongaussian}

The analysis described in the main text only includes the Gaussian contribution to galaxy stochasticity. In standard $\PplusB$ analysis, one usually includes a density-dependent stochastic contribution and a non-Gaussian white noise for the bispectrum, i.e. $\delta_{g}^{\mathrm{stoch.}}(\vx) = [b_\eps + b_{\eps\d} \d(\vx) +  b_{\nabla^2\eps} \nabla^2]\eps_0(\vx)+ b_{\eps^2}[\eps_0^2(\vx)-\langle \eps_0^2 \rangle]$, where $\eps_0$ is a unit Gaussian random field (see App.~A of \cite{Tucci:2023bag}). Although the density-dependent noise is more important than the contribution $\propto k^2$ \cite{Cabass:2019lqx}, it is technically more challenging to implement in FBI \cite{Cabass:2020nwf,Schmidt:2020tao}. In addition, it is usually the case that only the third-order bias operator that enters the power spectrum at one loop order \cite{Assassi:2014fva} is constrained, which corresponds to the operator $\Otd$ in our bias parameterization. 

\begin{figure}[htbp]
    \begin{minipage}[t]{0.5\linewidth}
        \centering
            \includegraphics[width=\linewidth]{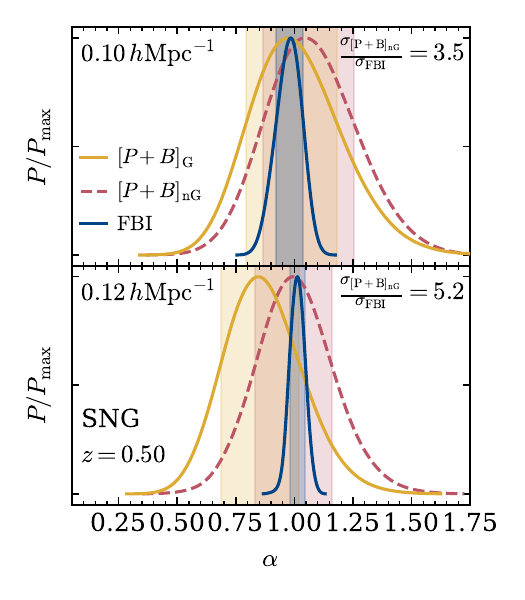}
        \label{fig:alpha_1D-constraint_FBI_vs_P+B-nG_SNGz05_2kmax}
    \end{minipage}\hfill
    \begin{minipage}[t]{0.5\linewidth}
        \centering
        \includegraphics[width=\linewidth]{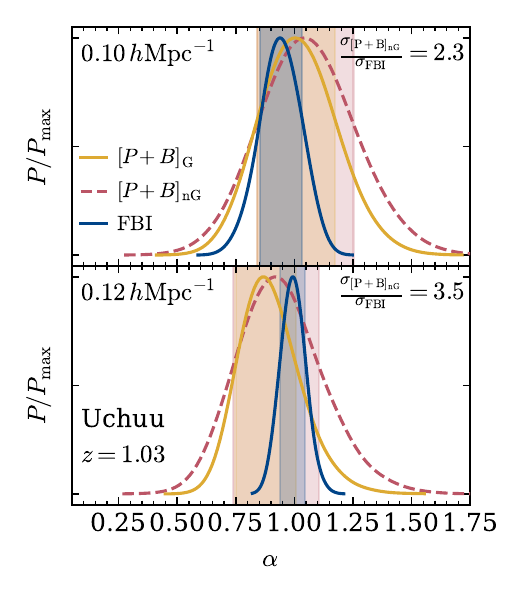}
        \label{fig:alpha_1D-constraint_FBI_vs_P+B-nG_Uchuuz10_2kmax}
    \end{minipage}
    \caption{Similar to Fig.~2 (left panels) and Fig.~3 (right panels) in the main text, but including the $\alpha$ posteriors for SBI $\PplusB$ analyses that include non-Gaussian (nG) contributions to galaxy stochasticity, as described in \cref{subsec:SBInongaussian}. Vertical bands indicate the 68\% limits of the posteriors. The ratios of the 1-$\sigma$ constraints between FBI (blue) and SBI $[\PplusB]_{\mathrm{nG}}$ (pink) are shown in the upper right corners.\label{fig:alpha_1D-constraint_FBI_vs_P+B-nG_2kmax}}
\end{figure}

We therefore consider an additional analysis where we fix all third-order bias parameters to zero with the exception of $b_{\mathrm{Otd}}$, which is inferred together with the two additional stochastic contributions $b_{\eps^2}$ and $b_{\eps\delta}$. We show in \cref{fig:alpha_1D-constraint_FBI_vs_P+B-nG_2kmax} the results of such analysis, which leads to similar constraints on $\alpha$ than the case with only Gaussian noise. We apply convergence and calibration tests also to these posteriors (not shown).

% End of SM

\end{document}